\documentclass{PoS}

\usepackage{amsmath}
\usepackage{caption}
\usepackage{subcaption}
\usepackage{float}

\title{EUSO-SPB1: Flight data classification and Air Shower Search Results}

\ShortTitle{EUSO-SPB1: data classification and Air shower search results}

\author{\speaker{Abraham D\'iaz Damian}, for the JEM-EUSO Collaboration\footnote{for collaboration list see PoS(ICRC2019)1177}\\
        Institut de Recherche en Astrophysique et Planetologie, Toulouse, France.\\
        E-mail: \email{abraham.diazdamian@irap.omp.eu}}

\abstract{The Extreme Universe Space Observatory on a Super Pressure Balloon (EUSO-SPB1) is the second balloon pathfinder of the JEM-EUSO collaboration. It is a nadir pointing UV telescope which aims at observing Ultra High Energy Cosmic Rays (UHECR) air showers through their fluorescence emission. It was launched the 24th of April, 2017, from the NASA balloon launch site in Wanaka, New Zealand. During it's flight, EUSO-SPB1 took data during 12 moonless nights until the termination of the mission. In this paper we present events found in triggered data while searching for air showers. We classify these events into different populations whose characteristics and origins we discuss. We show that the majority of our triggered events are direct Cosmic Ray hits on the detector. No air shower candidate have been found in this analysis.}

\FullConference{36th International Cosmic Ray Conference -ICRC2019-\\
		July 24th - August 1st, 2019\\
		Madison, WI, U.S.A.}

\begin{document}

\section{Introduction}
\label{introduction}

EUSO-SPB1 \cite{Wiencke2018} is the second balloon pathfinder mission of the JEM-EUSO program and the first one to incorporate an on-board First Level Trigger (FLT) logic \cite{Abdellaoui2017} to detect "from above" Extensive Air Showers (EAS) through their fluorescence emission. It was launched the 24th of April, 2017 from NASA's stratospheric balloon launch site located in Wanaka, New Zealand. EUSO-SPB1 performed observations for 12 moonless nights ($\approx$ 30 h) and returned about ninety thousand triggered event packets ($\approx$ 60 GB of data) before sinking in the Pacific ocean on May 6th 2019. This paper presents one analysis approach of the EUSO-SPB data. The data is visually inspected and categorized first in order to understand the triggered events, then an automated method is used to extract the observed features and create a catalog. The detected events are used to search for EAS candidates. A classification of the event population is presented and possible origins are discussed.

\section{EUSO-SPB detector and trigger logic}
\label{sec:instrument}

\begin{figure}
	\begin{center}
		\includegraphics[width=0.8\textwidth]{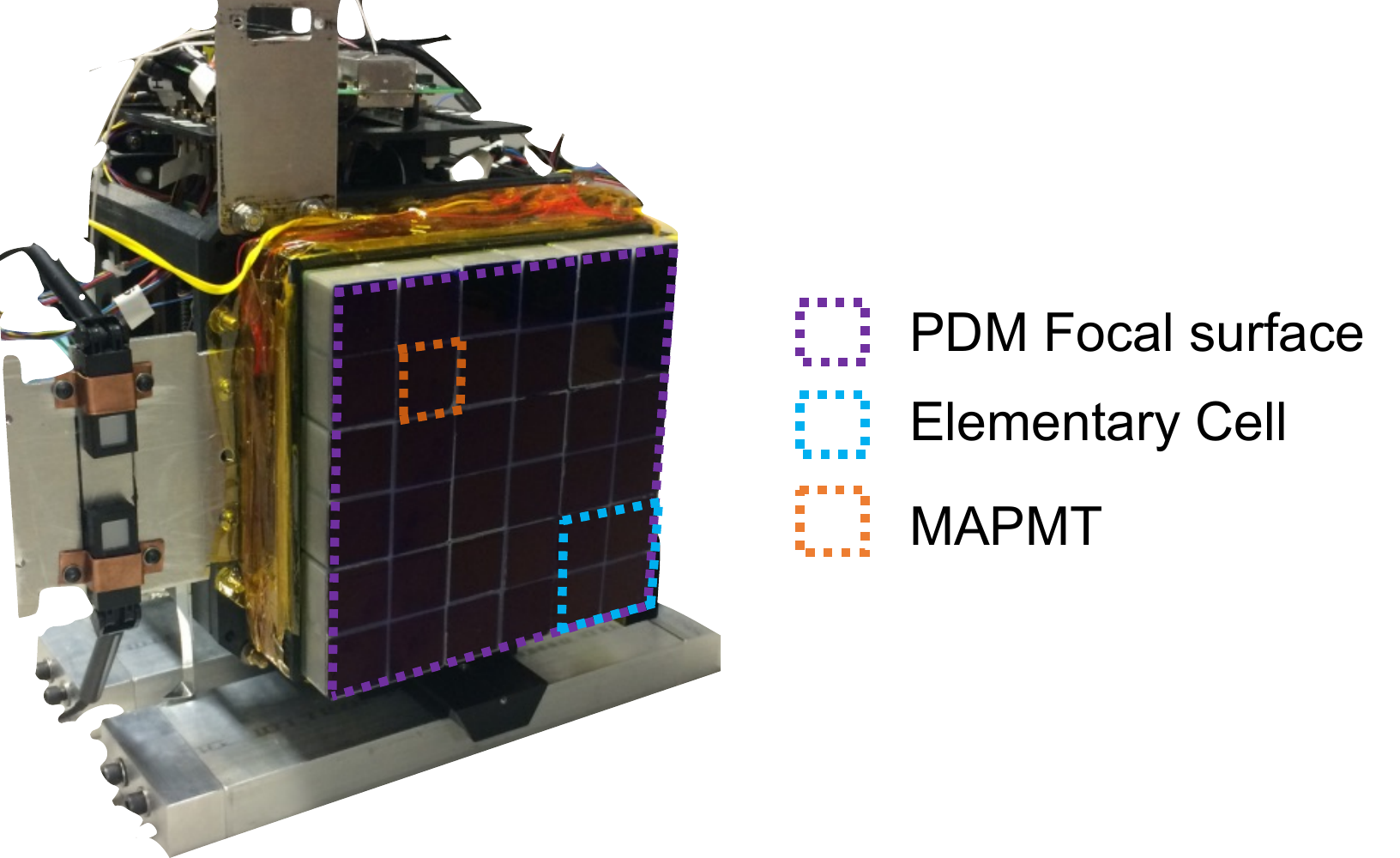}
	\end{center}
	\caption[Photo Detector Module]{Photo detector module of EUSO-SPB1. The focal surface, elementary cell and MAPMT are ensquared in violet, blue and orange respectively. To the left there are two UV Photodiodes used to monitor ambient light.}
	\label{fig:pdm}
\end{figure}

The detector of EUSO-SPB1 \cite{Bacholle2018} consists of one JEM-EUSO Photo detector module (PDM) which includes the focal surface (FS), FLT logic, high voltage power supply and associated electronics. The FS is composed of 36 (6$\times$6) Multi-Anode Photomomultipliers (MAPMTS) of 64 channels each (Hamamatsu R11265-113-M64). The MAPMTs are grouped for read-out purposes in Elementary Cells (EC) of 2$\times$2 MAPMTs and the PDM is composed of 9 ECs. The detector is operated in photon counting mode and has a time resolution of 2.5 $\mu$s called Gate Time Unit (GTU). Fig \ref{fig:pdm} shows the PDM of EUSO-SPB1.

The FLT operates at the EC level and is in charge of rejecting unwanted events and selecting EAS candidates.  It rejects UV night glow and electronic noise by requiring a persistent signal in a 3$\times$3 pixel cell above an average background threshold. If the threshold is exceeded on one cell the FLT checks if it happens again in a preset number (R) of GTUs within a (P) range of consecutive GTUs. If this condition is validated then a trigger is issued and the event is saved by assembling a packet of 128 continuous  GTUs (320 $\mu s$). During flight three different trigger configurations were used (P=0, R=0; P=1, R=1 and P=4, R=2). The trigger's flight configuration and performance are detailed in \cite{bertaina_trigger_2018, Battisti2019}. \\

\section{Data analysis method}
\label{sec:method}
\begin{figure}
    \centering
    \begin{subfigure}{0.3\textwidth}
        \includegraphics[width=\textwidth]{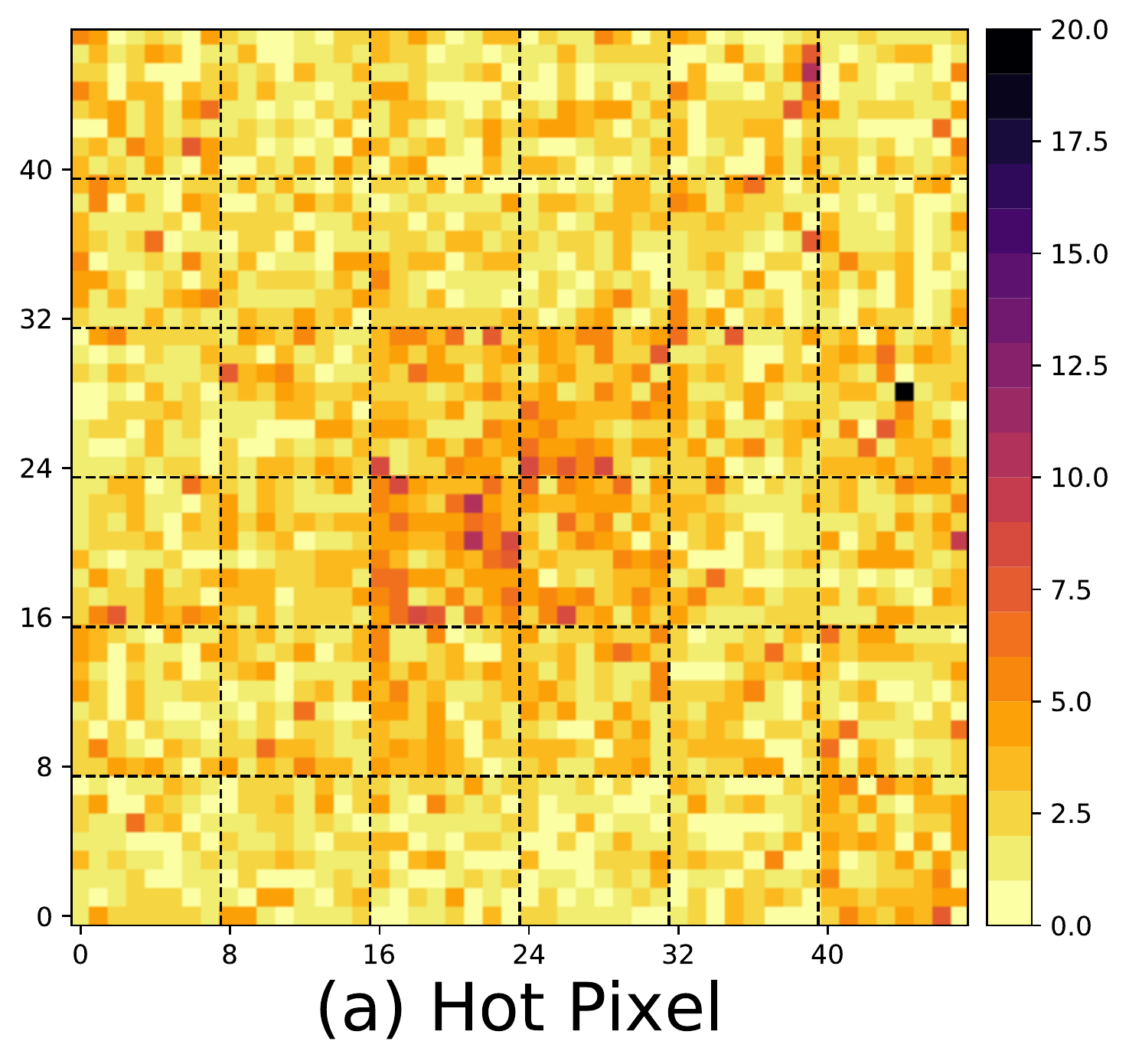}
    \label{fig:fig_pixel}
    \end{subfigure}
    \begin{subfigure}{0.3\textwidth}
        \includegraphics[width=\textwidth]{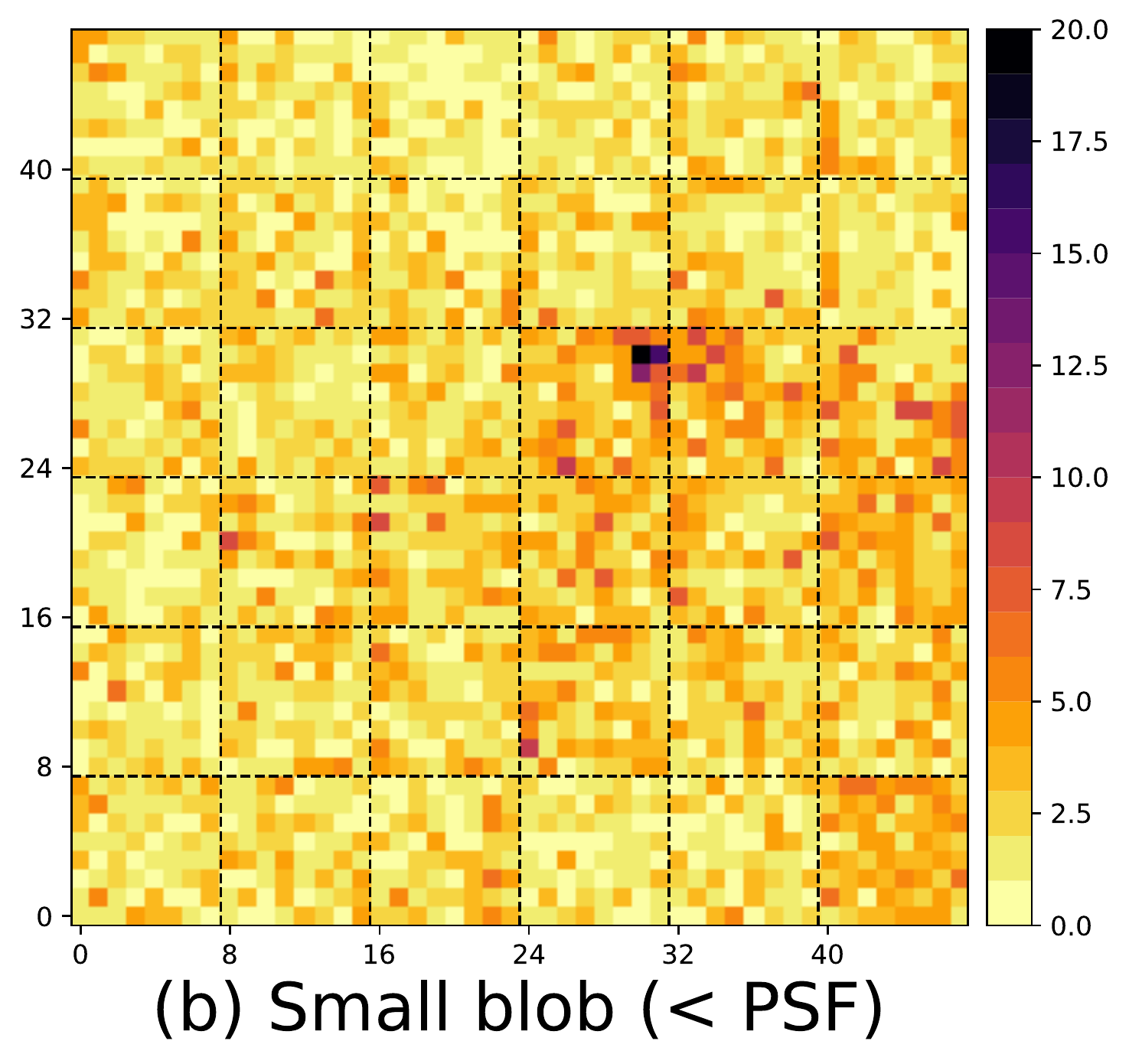}
        \label{fig:fig_small_blob}
    \end{subfigure}
    \begin{subfigure}{0.3\textwidth}
        \includegraphics[width=\textwidth]{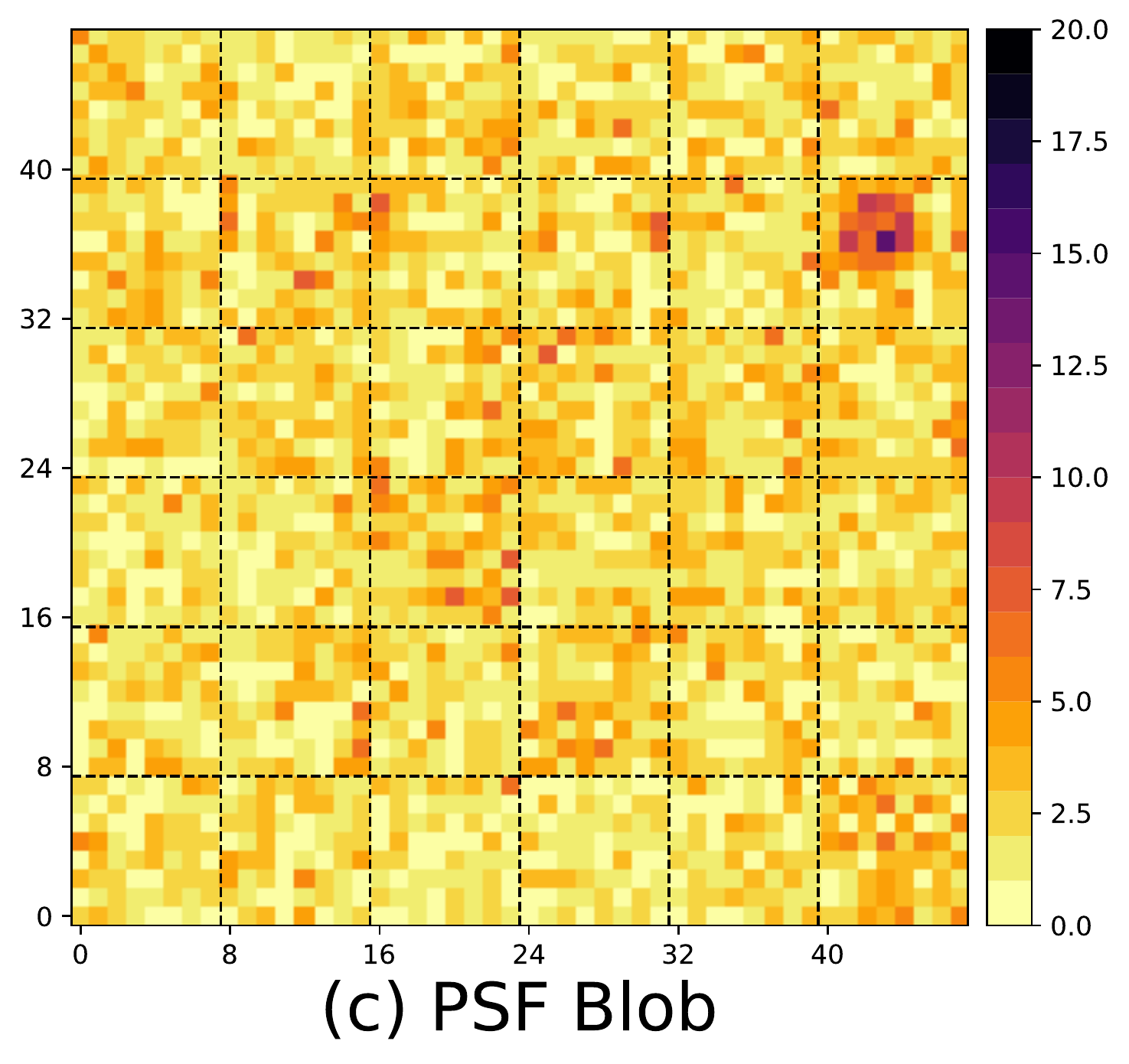}
        \label{fig:fig_psf_blob}
    \end{subfigure}
    \begin{subfigure}{0.3\textwidth}
        \includegraphics[width=\textwidth]{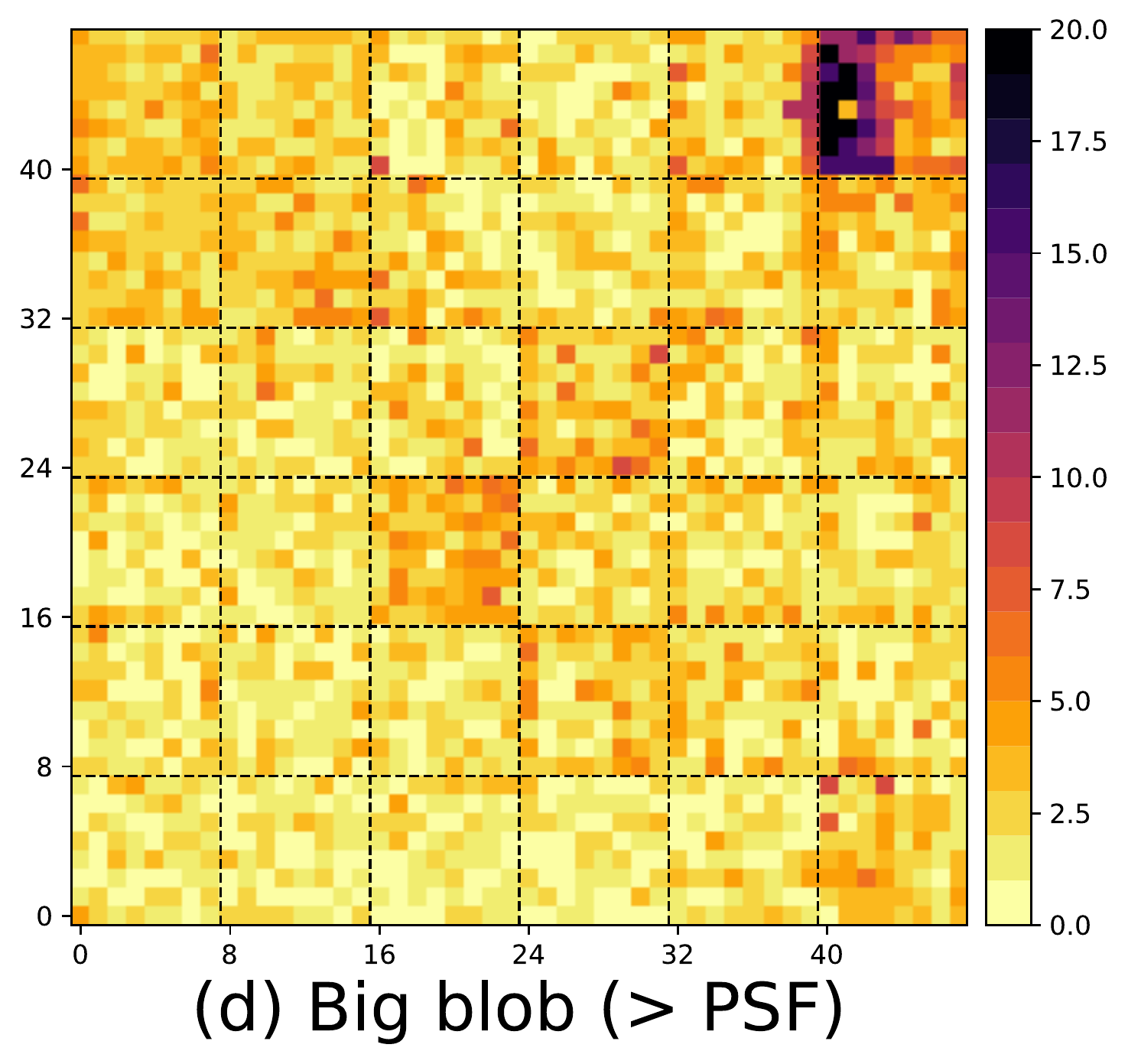}
        \label{fig:fig_big_blob}
    \end{subfigure}
    \centering
    \begin{subfigure}{0.3\textwidth}
        \includegraphics[width=\textwidth]{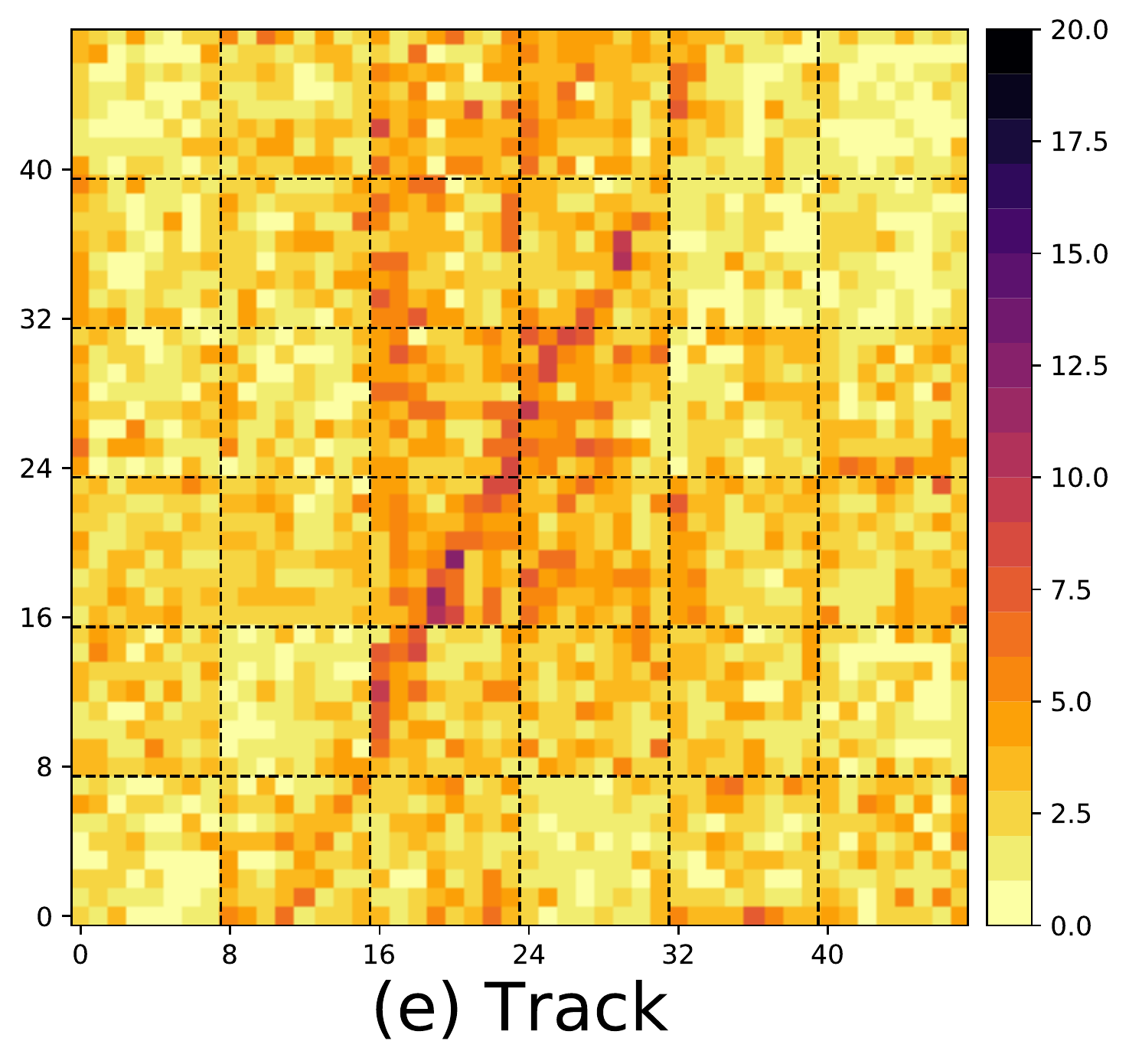}
        \label{fig:fig_track}
    \end{subfigure}
    \begin{subfigure}{0.3\textwidth}
        \includegraphics[width=\textwidth]{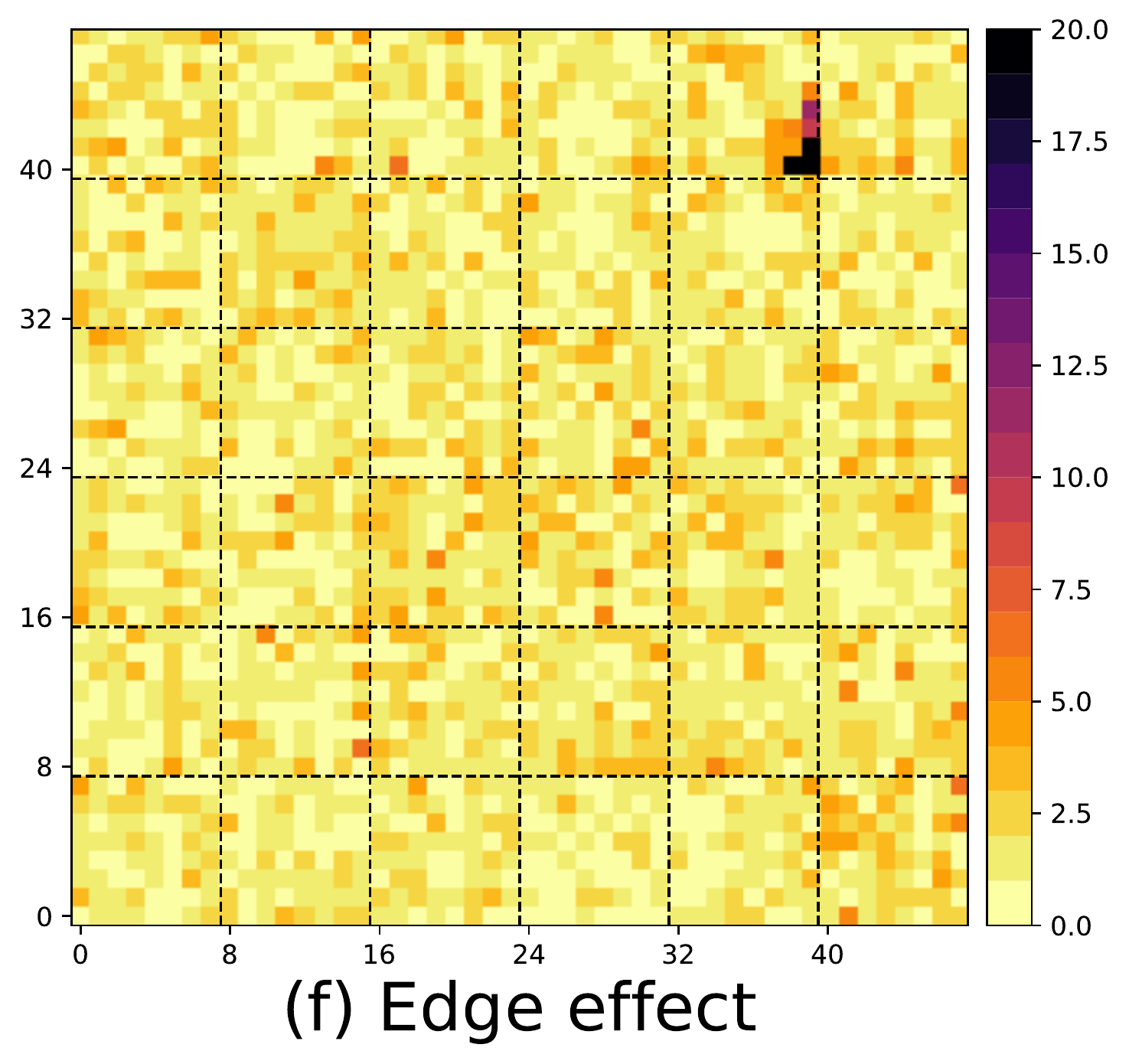}
        \label{fig:fig_edge_effect}
    \end{subfigure}
\label{fig:event_types}
\caption[Event categories]{Categories of observed events. From top left to bottom right: small blob, PSF blob, big blob, track, hot pixel, edge effect. The black dotted lines denote the MAPMT borders, in reality there is a gap between each one}
\label{fig:event_categories}
\end{figure}

To understand better the events triggered during flight the first approach taken was to visually inspect the data. Eye observation showed that the events consisted in pixel excesses of different sizes and shapes. The events are typically 1 GTU long, however in some cases the signal persists longer. These pixel excesses include single pixels, pixel groups of varying sizes called "blobs" and linear "track" features. The blobs were categorized as a function of their size with respect to the point spread function (PSF) of the instrument (3x3 pixels). Additionally, many of these events were frequently located and constrained in the edge pixels of an MAPMT and because of this they were considered as a category of their own. In total the following categories of events were identified: (a) hot pixel, (b) small blob (< PSF), (c) PSF blob, (d) big blob (> PSF),  (e) track and (f) edge effect. Figure \ref{fig:event_categories} shows one frame of each type of event.

\begin{figure}
	\begin{center}
		\includegraphics[width=0.99\textwidth]{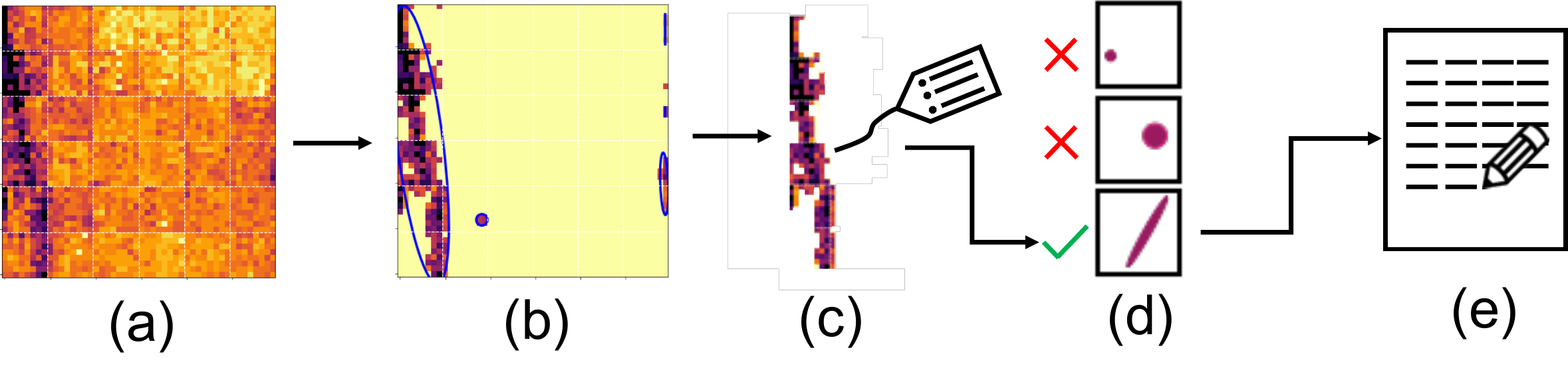}
	\end{center}
	\caption[GTU frame processing method]{GTU frame processing method: (a) The raw GTU frame is processed. (b) Background pixels are removed, neighboring pixels are grouped into regions and the region is fitted with an ellipse. (c) The properties of the region are extracted. (d) The region is classified and (e) saved into a data set for further analysis.}
	\label{fig:method}
\end{figure}

Following a first phase of visual inspection an automated method was developed to extract the observed features. Data is processed on a GTU basis by extracting the foreground features from the background as follows: the average background per pixel per packet is subtracted. The image is then filtered in the spatial frequency domain to remove high frequency noise. To remove the background and isolate significant pixels a threshold is set using Otsu's method \cite{Otsu1979}. The remaining pixels are grouped together into "regions" if they're within two steps in the grid from each other. Each region is fitted with an ellipse, the region properties are extracted and it is classified. The classification is based on the pixel size of the region and the eccentricity of the fitted ellipse. More eccentric features are more likely to be tracks. In some cases, after subtracting the background there is more than one leftover region, specially for noisier images. To keep one event per frame the region with the highest pixel value is kept and saved in a dataset for further analysis. In order to look for EAS, continuous GTU events are stitched together and analysed. The data processing method is schematically explained in fig. \ref{fig:method}

\section{Results}
\label{sec:results}

\subsection{Search for extensive air showers}
\begin{figure}
    \centering
    \begin{subfigure}{0.38\textwidth}
        \includegraphics[width=\textwidth]{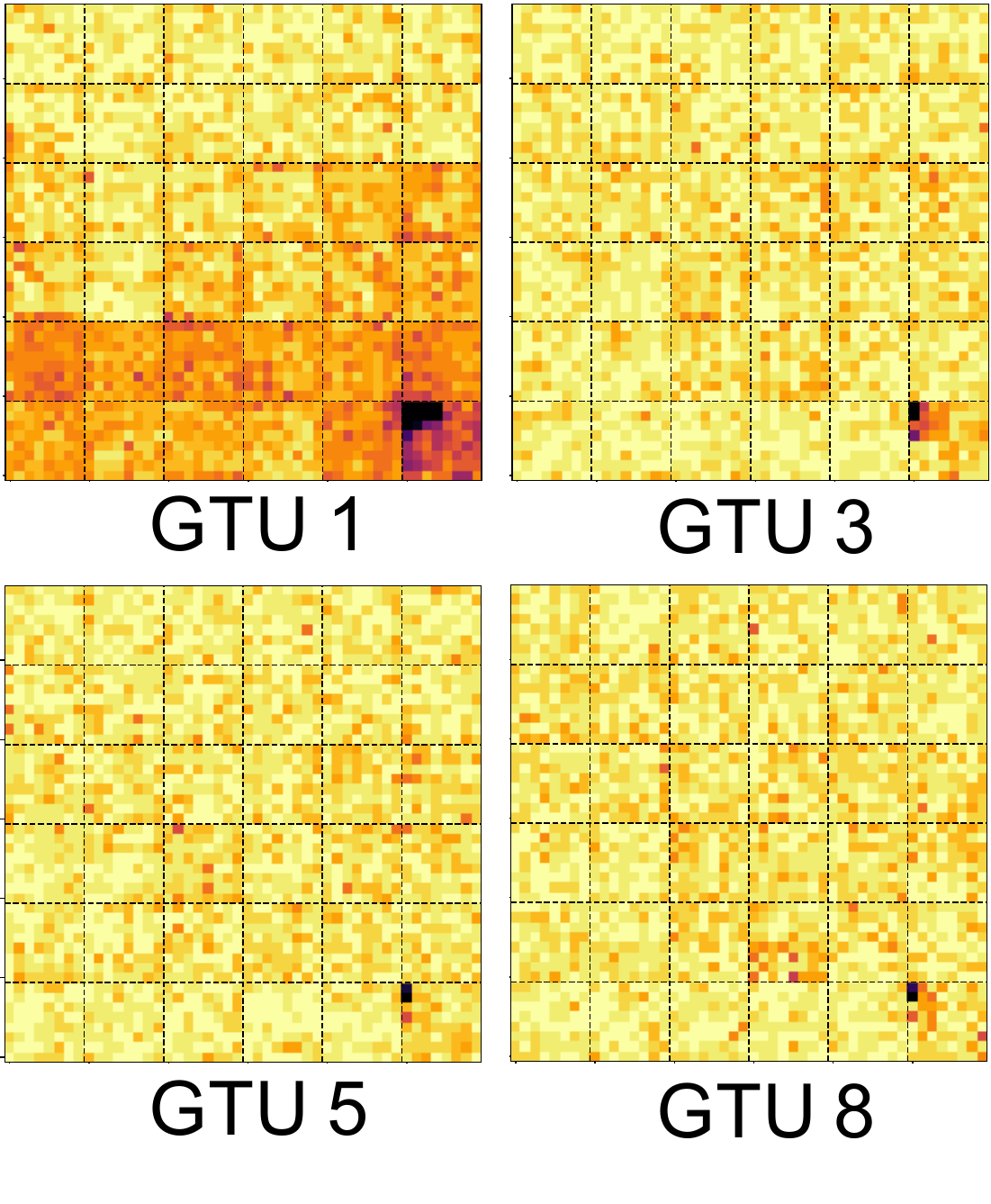}
        \caption{}
        \label{fig:8 GTU event}
    \end{subfigure}
    \begin{subfigure}{0.55\textwidth}
        \includegraphics[width=\textwidth]{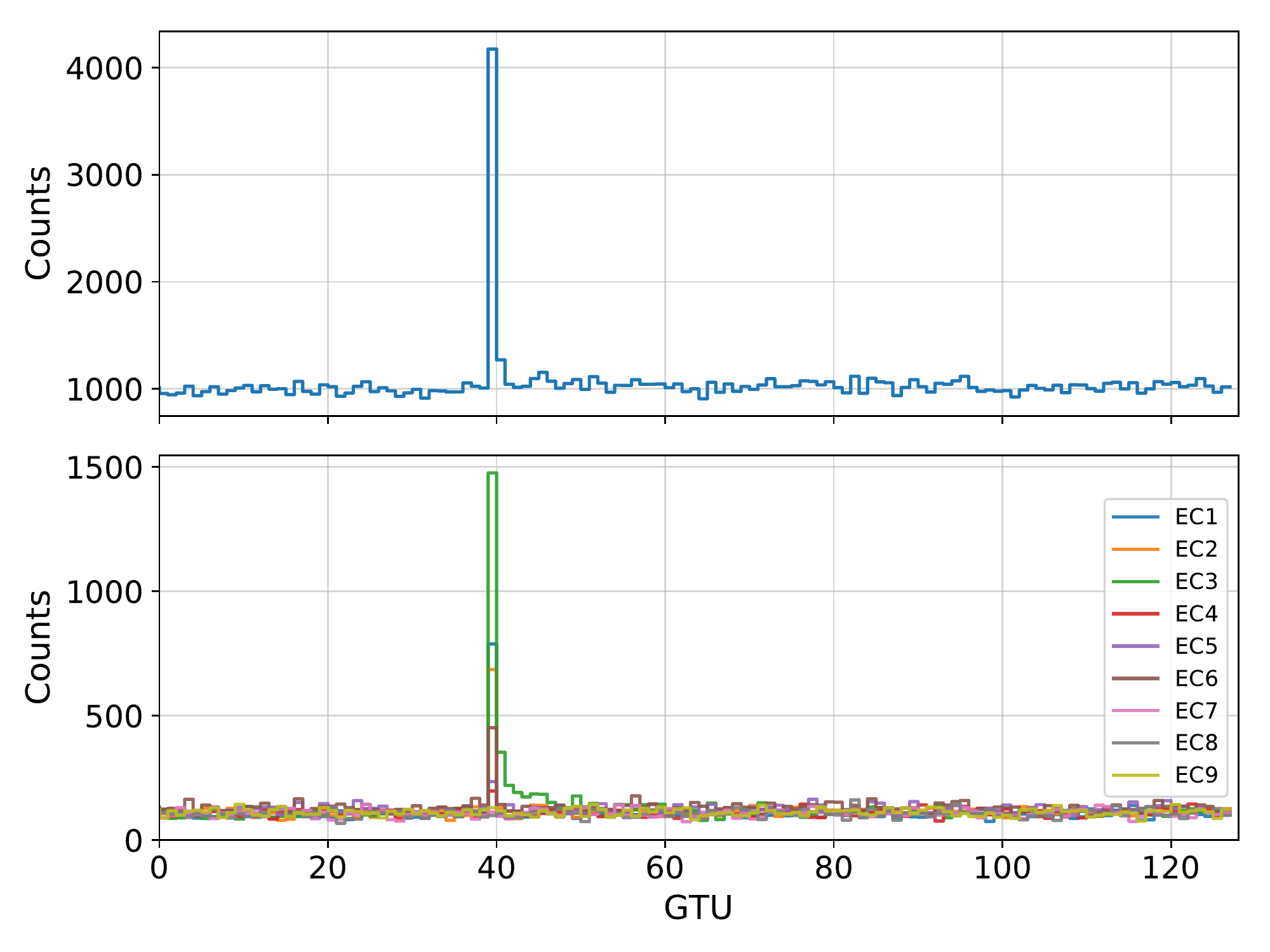}
        \caption{}
        \label{fig:event_counts}
    \end{subfigure}
\caption[Long Event]{Image and time signal of an 8 GTU long event. (a) 4 frames of the event. (b) Counts per GTU at the PDM (top) and EC (bottom) level where a decaying signal tail is visible. }
\label{fig:long_event}
\end{figure}

The flight data was reduced down to about 89k events of interest for further analysis and visual inspection. EAS candidates were searched for by joining consecutive events longer than 3 GTU. This produced 4128 multi-GTU events ranging from 3 to 40 GTU long (16236 GTU in total). Fig \ref{fig:long_event} shows the frames from one of these events and the time signal at the EC and PDM level. All multi-GTU events have been visually inspected and none showed EAS signatures. These events follow a recurring pattern: they are static and localized in a single MAPMT. They start as a blob which then decays into a single pixel or edge effect. For an EAS there should be a constant sized blob propagating across the detector each GTU.

\subsection{Population of events}

\begin{figure}
    \centering
    \begin{subfigure}{0.49\textwidth}
        \includegraphics[width=\textwidth]{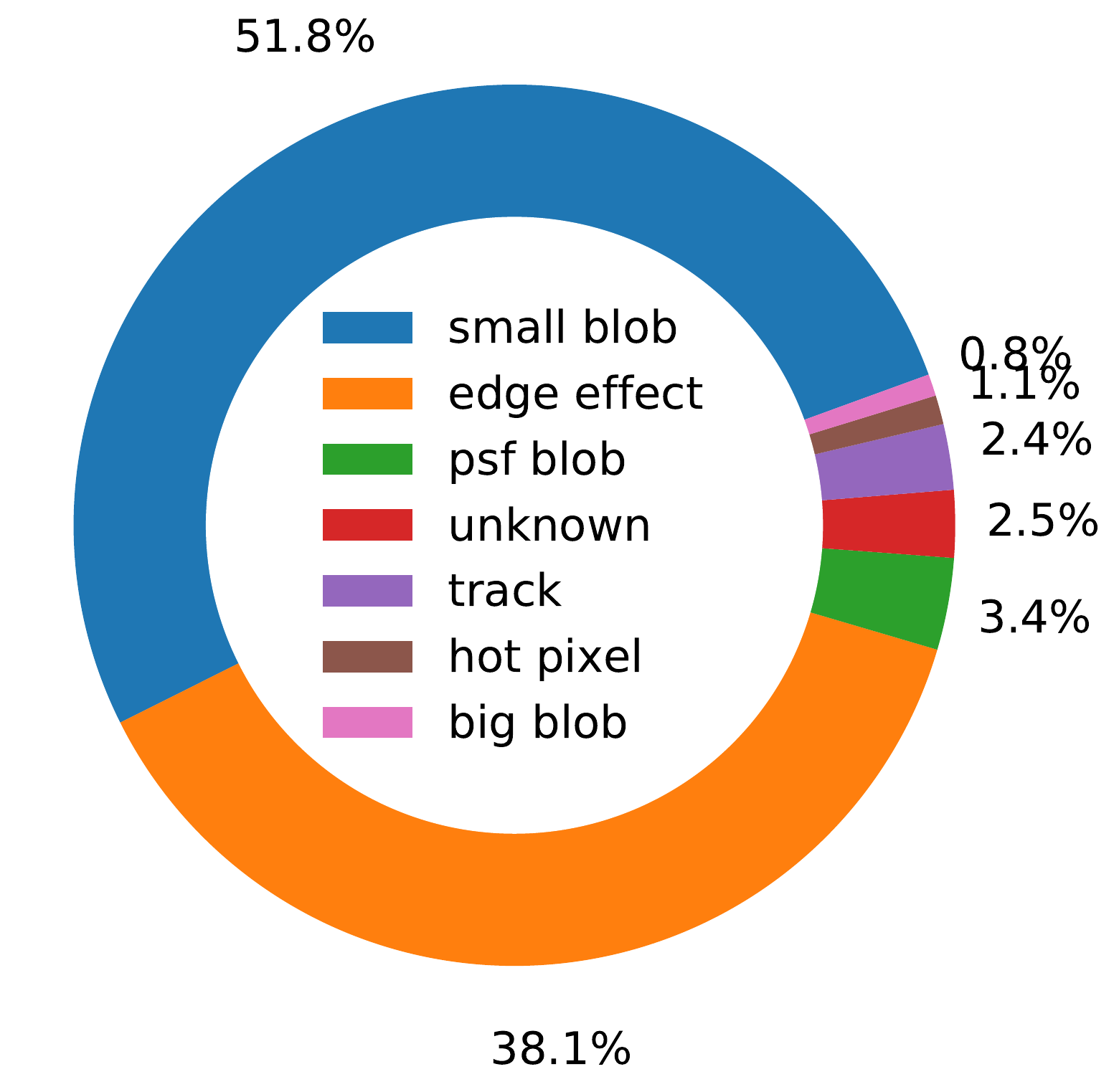}
        \caption{}
        \label{fig:population}
    \end{subfigure}
    \begin{subfigure}{0.49\textwidth}
        \includegraphics[width=\textwidth]{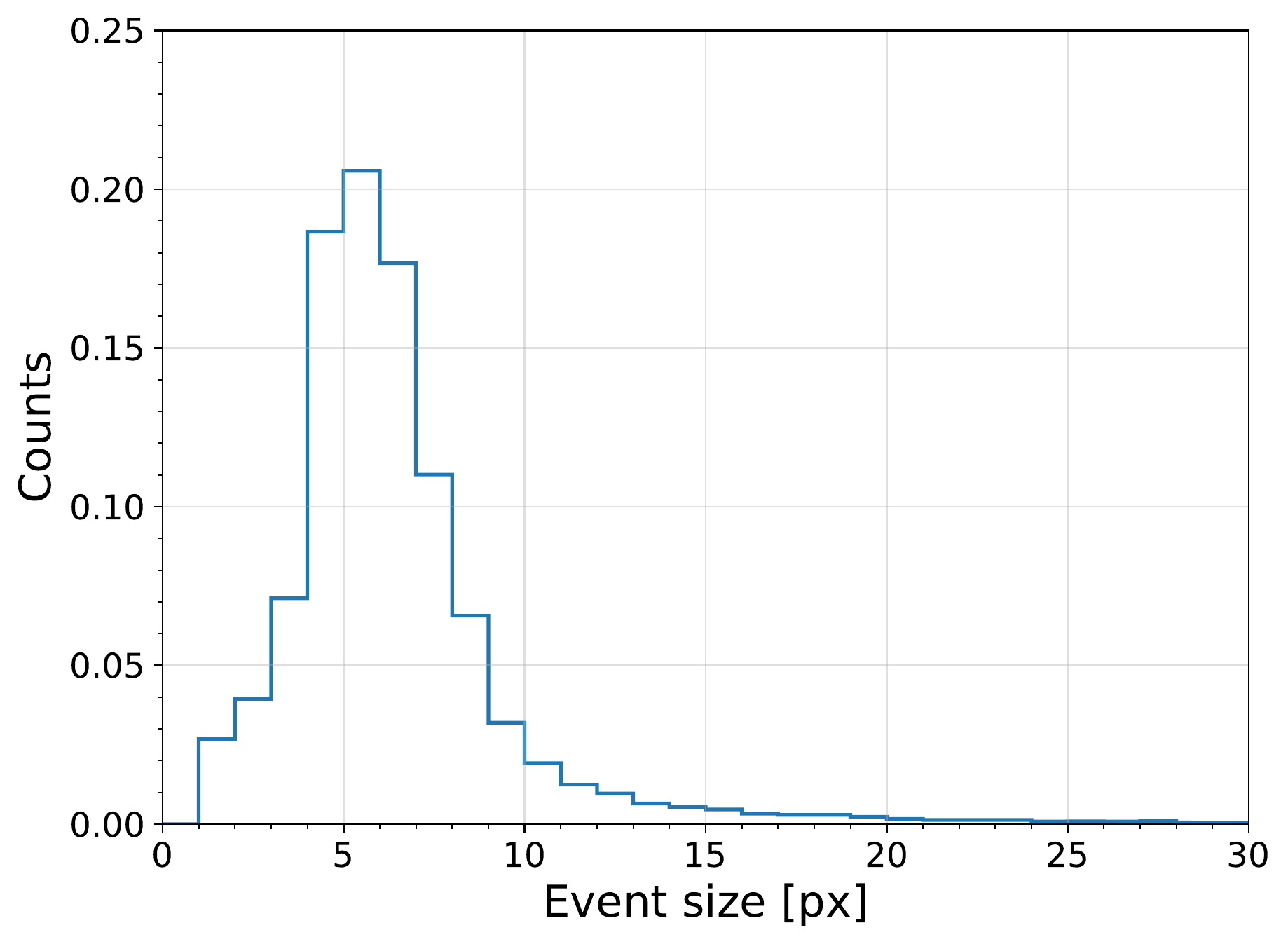}
        \caption{}
        \label{fig:event_size}
    \end{subfigure}

\caption[Event population and size of events]{Population and size of observed events. (a) Small blobs and edge effect are the most common events comprising about 90\% of the total, the unknown category is for events which didn't fit the classification parameters. (b) The most common event size is 5 pixels.}
\label{fig:event_population_size}
\end{figure}

The statistics of the event population in figure \ref{fig:population} shows that most of the detected events are small sized. Small blobs (51.8\%), edge effects (38.1\%) and hot pixel events (1.1\%) account for 91\% of all events. PSF blobs Events that did not fit into any category are classified as unknown (2.5\%). Figure \ref{fig:event_size} shows the distribution of event pixel size, the most common event size has a spread of 5 pixels which is smaller than the expected PSF size. The size and short length of the majority of detection, point at non-EAS events since it would take a few GTU for an EAS to cross the detector's field of view. Understanding these events is therefore crucial for the improvement of future triggers.

\begin{figure}
	\begin{center}
		\includegraphics[width=0.75\textwidth]{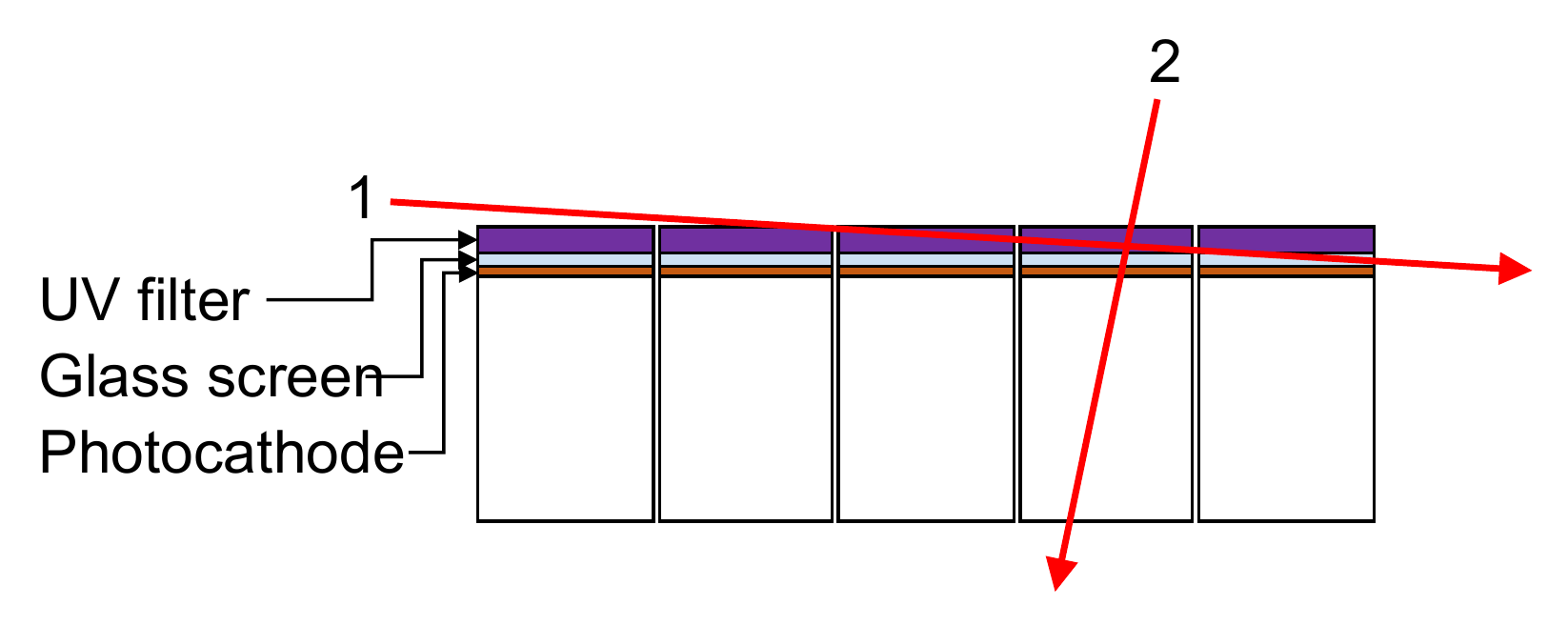}
	\end{center}
	\caption[Cosmic Ray Hits]{Cosmic Ray Hits on a Photomultiplier array detector. In scenario 1 the cosmic ray arrives at an angle approximately parallel to the Focal Surface projecting forming linear pattern. In scenario 2 a cosmic ray hits at an almost perpendicular incidence angle doing a more punctual interaction. Scenario 2 is the presumed origin of blobs.}
	\label{fig:cr_hits}
\end{figure}

Blobs and tracks are presumed to be caused by direct cosmic ray hits in the detector, given the environment of the balloon-craft and the linear pattern of tracks. Similar track events have also been observed by the TUS space observatory \cite{Khrenov2017} whose detector is an array of Photomultiplier tubes (PMT). It has been shown by simulations that tracks are produced by protons that hit the detector's UV glass filter at angles almost parallel to their plane. While traversing the glass, the protons produce fluorescence and Cherenkov photons which illuminate pixels in a linear pattern and induce triggers. Extrapolating this scenario to cosmic rays hits at oblique incidence angles, a more punctual interaction is produced, creating blobs smaller than the PSF (see figure \ref{fig:cr_hits}). Simulations are underway to test this scenario on the EUSO-SPB1 detector geometry.

Edge effects are attributed to instrumental causes, given their linear pattern and frequent placement along the MAPMT borders. This is presumably due to an inhomogeneity in the MAPMT construction which favors two of its edges with higher gains. Further study is being conducted on this effect.

Despite the absence of EAS candidates, the catalogue of extracted events has proven useful to understand the triggered data and will help to test new triggers in the aims of rejecting anomalous and background events.

\section{Conclusion}
\label{sec:conclusion}

The triggered data of the EUSO-SPB1 experiment has been analysed using a mixed approach of visual inspection and automated method to extract significant features observed in the data. EAS events were searched by looking at 4128 events longer than 3 GTU frames, however no candidate has been found.

The triggered events have been shown to be pixel blobs of different sizes, linear features along photomultiplier tube edges and tracks spanning multiple pixels and MAPMT. The most common events are small blobs (< 9 px) with 51.8 \% of the total. These events, as well as tracks are caused by direct cosmic ray hits on the detector which creates a signal excess typically lasting 1 GTU and inducing a trigger. Understanding these events is important for the improvement of future missions and works are underway to design more robust triggers which suppress unwanted events.

\section{Acknowledgements}
This work was partially supported by Basic Science Interdisciplinary Research Projects of RIKEN and JSPS KAKENHI Grant (22340063, 23340081, and 24244042), by the Italian Ministry
of Foreign Affairs and International Cooperation, by the Italian Space Agency through the ASI INFN agreement n. 2017-8-H.0, by NASA award 11-APRA-0058 in the USA, by the Deutsches Zentrum für Luft- und Raumfahrt, by the Helmholtz Alliance for Astroparticle Physics funded
by the Initiative and Networking Fund of the Helmholtz Association (Germany), and by Slovak Academy of Sciences MVTS JEM-EUSO as well as VEGA grant agency project 2/0132/17. We acknowledge support from French space agency CNES. Russia is supported by the Russian Foun- dation for Basic Research Grant No 16-29-13065-ofi-m. The Spanish Consortium involved in JEM- EUSO is funded by MICINN \& MINECO under the Space Program projects: AYA2009-06037- E/AYA, AYA-ESP2010-19082, AYA-ESP2011-29489-C03, AYA-ESP2012-39115-C03, AYA- ESP 2013 - 47816-C4, MINECO/FEDER-UNAH13-4E-2741, CSD2009-00064 (Consolider MULTIDARK)
and by Comunidad de Madrid (CAM) under projects S2009/ESP-1496 \& S2013/ICE-2822.

\bibliographystyle{JHEP}
\bibliography{references.bib}

\end{document}